\documentclass[12pt,a4paper]{article}

\usepackage{a4wide}
\usepackage{amsmath,amssymb}
\usepackage{graphicx}
\usepackage{color}

\usepackage{hyperref}

\def\SU#1{{SU}(#1)}
\def\U#1{{U}(#1)}
\def\dg{\dagger}
\def\doublet#1#2{\begin{pmatrix}#1\\#2 \end{pmatrix}}
\def\M{\mathcal M}
\def\Re{\mathrm{Re}}
\def\Im{\mathrm{Im}}
\def\imag{i}
\def\e{\mathrm{e}} 
\def\s{\mathrm{s}}
\def\c{\mathrm{c}}
\def\CP{\textit{CP}\/}
\def\GeV{\textrm{ GeV}}
\def\m{\scriptstyle}

\title{Tree-unitarity bounds for THDM Higgs masses revisited}
\author{J. Ho\v rej\v s\'\i, M. Kladiva\\
Institute of Particle and Nuclear Physics, Faculty of Mathematics and
Physics,\\
Charles University, V Hole\v sovi\v ck\' ach 2, CZ-180 00 Prague 8, Czech Republic }
\date{October 12, 2005} 

\begin{document}
\maketitle

\begin{abstract}
We have reconsidered theoretical upper bounds on the scalar boson masses
within the two-Higgs-doublet model (THDM), employing the well-known
technical condition of tree-level unitarity. Our treatment provides a
 modest
extension and generalization of some previous results of other authors. We
present a rather detailed discussion of the solution of the relevant
inequalities and offer some new analytic formulae as well as numerical
values for the Higgs mass bounds in question. A comparison is made with the
earlier results on the subject that can be found in the literature.
\end{abstract}

\section{Introduction}
 The two-Higgs-doublet model (THDM) of electroweak interactions is one of the
simplest extensions of the Standard Model (SM). It incorporates two complex
scalar doublets in the Higgs sector, but otherwise its structure is the same as
that of the SM. Obviously, such a theory is rather appealing on purely
aesthetic grounds: in view of the familiar doublet pattern of the elementary
fermion spectrum, one can speculate that an analogous organizational principle
might work for the "scalar Higgs matter" as well. Further, any Higgs sector
built upon doublets only is known to preserve naturally the famous lowest-order
electroweak relation $\rho=1$ (where $\rho=m_W^2/(m_Z^2 \cos^2\theta_W)$), which
has been tested with good accuracy.  On the phenomenological side, an important
aspect of the THDM is that its Higgs sector may provide an additional source of
\CP\ violation; in fact, this was the primary motivation for introducing such a
model in the early literature on spontaneously broken gauge theories in
particle physics \cite{Lee:1973}. Of course, there is at least one more reason
why the THDM has become popular\footnote{For useful reviews of the subject see
e.g. \cite{Sher}, \cite{Guide}} during the last two decades or so: its Higgs
sector essentially coincides with that of the minimal supersymmetric SM (MSSM),
but the values of the relevant parameters are less restricted. The spectrum of
physical Higgs particles within THDM consists of five scalar bosons, three of
them being electrically neutral (denoted usually as $h$, $H$ and $A^0$) and
the other two charged ($H^\pm$).  At present, some partial information
concerning direct experimental lower bounds for the Higgs masses is available,
coming mostly from the LEP data (cf. \cite{Experiment}).

  On the other hand, it is also interesting to know what could be possible
theoretical limitations for masses of the so far elusive Higgs particles within
such a "quasi-realistic" model. For this purpose, some rather general methods
have been invented, based mostly on the requirements of internal consistency of
the quantum field theoretical description of the relevant physical quantities.
One particular approach, which is perhaps most straightforward in this regard,
relies on perturbative unitarity of the $S$-matrix. In its simplest form it is
implemented at the lowest order, by imposing unitarity constraints on the
tree-level amplitudes of a suitable set of scattering processes. Let us recall
that this technique was originally developed by B.W. Lee, C. Quigg and H. Thacker (LQT),
who employed it in their well-known analysis of perturbative upper bound for
the SM Higgs boson mass~\cite{LQT}. The LQT method was subsequently applied also to
electroweak models with extended Higgs sectors; some results can be found under
refs. \cite{OldPapers},
\cite{KKT}, \cite{AAN}. In particular, authors of the papers \cite{KKT}, \cite{AAN}
 analyzed in this way a restricted version of the THDM with \CP-conserving Higgs
sector and obtained slightly differing values of the bounds in question (due to
slightly different implementations of the LQT method). Recently, the issue of
tree-unitarity constraints for THDM Higgs boson masses has been taken up
again in the work \cite{Ginzburg:2003} (see also~\cite{Ginzburg:2004},\cite{Ginzburg:2005}), where
a rather general model involving \CP\ violation has been considered;
this seems to be another vindication of the persisting interest in the subject.

  The purpose of the present paper is to supplement and extend the existing
results concerning the THDM Higgs mass upper bounds. We carry out a rather
detailed analysis of a relevant set of inequalities that follow from the
requirement of tree-level unitarity. In particular, the procedure of explicit
solution of these constraints is discussed in considerable detail and, among
other things, some results of the corresponding numerical calculations within a
general THDM are presented. For the model without \CP\ violation we were able to
find a set of analytic expressions as well. Note that in this latter case, most
of the calculational details are contained also in an earlier unpublished work
by one of us (see~\cite{diplomka}). Let us also remark that
there is no substantial overlap of the material presented in
\cite{Ginzburg:2003,Ginzburg:2004,Ginzburg:2005} with our results, so
 we believe that it makes sense to offer our
detailed analysis as a contribution to the current literature on the
particular problem in question.

  The plan of our paper is as follows: In Sect. \ref{sec:potential} the THDM
scalar potential and the scalar fields are described in some detail, 
in Sect. \ref{sec:LQT} we summarize briefly the LQT method and its
implementation within THDM and in Sect.
\ref{sec:inequalities} the relevant inequalities
expressing the tree-unitarity constraints are examined. The main analytic
results for the mass bounds in question are contained in sections
\ref{sec:MaMpm}, \ref{sec:MhMH}, \ref{sec:Mlightest} and Sect. \ref{sec:numeric}
contains numerical results obtained in the \CP-violating case (where we have not
been able to find analytical results).  The main results are summarized in Sect.
\ref{sec:conclusion}.
 
\section{THDM scalar potential}
 \label{sec:potential}

The most general scalar potential within THDM that is invariant under
$\SU2\times\U1$ can be written as (cf. \cite{Georgi} or \cite{Guide})
\begin{multline}
 V(\Phi)=\lambda_1 \left( \Phi_1^\dg \Phi_1 - \tfrac{v_1^2}2 \right)^2 +
         \lambda_2 \left( \Phi_2^\dg \Phi_2 - \tfrac{v_2^2}2 \right)^2 +
         \lambda_3 \left( \Phi_1^\dg \Phi_1 - \tfrac{v_1^2}2 +
                          \Phi_2^\dg \Phi_2 - \tfrac{v_2^2}2 \right)^2 +
\\
         \lambda_4 \left[ (\Phi_1^\dg \Phi_1) (\Phi_2^\dg \Phi_2) - 
                          (\Phi_1^\dg \Phi_2) (\Phi_2^\dg \Phi_1) \right] +
         \lambda_5 \left[ \Re(\Phi_1^\dg \Phi_2) - \tfrac{v_1 v_2}2 \cos\xi \right]^2 +
         \lambda_6 \left[ \Im(\Phi_1^\dg \Phi_2) - \tfrac{v_1 v_2}2 \sin\xi \right]^2
 \label{eq:potential}
\end{multline}
Note that such a form involves \CP\ violation, which is due to 
 $\xi\neq0$ \cite{Guide}.
 It also possesses an approximate discrete $Z_2$ symmetry under
 $\Phi_2 \to -\Phi_2$; this is broken "softly",
 by means of  the quadratic term
 \begin{multline}
   v_1 v_2 \left( \lambda_5 \cos\xi\, \Re(\Phi_1^\dg \Phi_2) +  \lambda_6 \sin\xi\,\Im(\Phi_1^\dg \Phi_2) \right) 
   = v_1 v_2\,\Re\left[ \left( \lambda_5 \cos\xi - \imag \lambda_6 \sin\xi \right) \Phi_1^\dg \Phi_2 \right]
 \end{multline}
  Let us recall that the main purpose of such an extra partial symmetry within
THDM is to suppress naturally the flavour-changing processes mediated by
neutral scalar exchanges that could otherwise arise within the quark Yukawa
sector \cite{Glashow:1976}. Note also that if such a symmetry were
exact, there would be no \CP\ violation in the Higgs sector of the considered
model.  For further remarks concerning the role of the $Z_2$ symmetry see 
e.g.~\cite{Ginzburg:2003} and references therein. As a quantitative measure of
the $Z_2$ violation we introduce  a parameter $\nu$, defined as
 \begin{equation}
   \nu=\sqrt{\lambda_5^2 \cos^2\xi + \lambda_6^2 \sin^2\xi}
   \label{eq:nu}
  \end{equation}
 (note that our definition of the $\nu$ differs slightly 
from that used in \cite{Ginzburg:2003}.)
  The minimum of the potential \eqref{eq:potential} occurs at
  \begin{equation}
    \Phi_1= \frac1{\sqrt2}\doublet0{v_1}, \qquad
    \Phi_2= \frac1{\sqrt2}\doublet0{v_2} \e^{\imag\xi}
  \end{equation}
  where we have adopted, for convenience, the usual simple choice of phases.
 Such a minimum determines vector boson masses through the Higgs mechanism; in
 particular, for the charged $W$ boson one gets $m_W^2=\frac12g^2 (v_1^2 +
 v_2^2)$, with $g$ standing for $\SU2$ coupling constant.  In a standard
 notation one then writes  $v_1=v\cos\beta, v_2=v\sin\beta$, where 
 $v$ is the familiar electroweak scale, $v=(G_F \sqrt{2})^{-1/2}\doteq 246 \GeV$
  and $\beta$ is a free parameter.
 THDM involves eight independent scalar fields: three of them can be identified
 with the would-be Goldstone bosons $w^\pm, z$ (the labelling is chosen so as to indicate
 that they are direct counterparts of the massive vector bosons $W^\pm, Z$
 within an $R$-gauge) and the remaining five correspond to physical Higgs
 particles --- the charged $H^\pm$ and the neutral ones $h, H, A^0$.

 We will now describe the above-mentioned Goldstone and Higgs bosons in more
 detail. To this end, let us start with a simple representation of  the
 doublets, namely  
 \begin{equation}
   \Phi_1=\doublet{w_1^-}{\frac1{\sqrt{2}}(v_1 + h_1 + \imag z_1 )}
    \qquad
   \Phi_2=\doublet{w_2^-}{\frac1{\sqrt{2}}(\e^{\imag\xi}v_2 + h_2 + \imag z_2 )}
   \label{eq:param}
  \end{equation}
 Of course, the scalar fields introduced in \eqref{eq:param} are in general unphysical; the
$w_{1,2}^\pm$ are taken to be complex and the remaining ones real, but
otherwise arbitrary. Note that an advantage of such a
parametrization is that the form of the quartic interactions is then the 
same as in \CP-conserving case.  The proper Goldstone and
Higgs fields are found through a~diagonalization of the quadratic part of the
potential \eqref{eq:potential}. When doing it, a convenient starting point is 
a~slightly
modified doublet parametrization
  \begin{equation}
    \Phi_1=\doublet{w_1^-}{\frac1{\sqrt{2}}(v_1 + h_1 + \imag z_1 )}
    \qquad
    \Phi_2=\doublet{w_2^{\prime-}}{\frac1{\sqrt{2}}(v_2 + h'_2 + \imag z'_2 )}
    \e^{\imag\xi}
     \label{eq:inaparam}
   \end{equation}
  that is obtained from \eqref{eq:param} by means of the unitary transformation $h'_2=h_2\cos\xi + z_2\sin\xi$,
  $z'_2=z_2\cos\xi-h_2\sin\xi$ a $w_2^{\prime\pm}=\e^{-\imag\xi}w_2^\pm$ . 
  Next, the scalar fields in \eqref{eq:inaparam} are rotated pairwise as
  \begin{gather}
    \doublet{H'}{h'} = 
     \begin{pmatrix} 
        \cos\beta & \sin\beta \\ -\sin\beta & \cos\beta 
     \end{pmatrix} 
     \doublet{h_1}{h'_2}
    \qquad
    \doublet{A'}{z} = 
      \begin{pmatrix} 
        \cos\beta & \sin\beta \\ -\sin\beta & \cos\beta 
      \end{pmatrix} 
      \doublet{z_1}{z'_2}
   \\
    \doublet{\zeta}{w} = 
      \begin{pmatrix} 
         \cos\beta & \sin\beta \\ -\sin\beta & \cos\beta 
      \end{pmatrix} 
     \doublet{w_1}{w'_2}
   \end{gather}
 When the quadratic part of \eqref{eq:potential} is recast in terms of the new
 variables, one finds out that the $z,w^\pm$ are massless Goldstone
 bosons and the $H^\pm$ represent massive charged scalars. At this stage, the
 fields $h',H' ,A'$ are still mixed and their mass matrix reads
 \begin{equation}
 \frac12
 \left(
 \begin{smallmatrix}
       \s_{2\beta}^2 (\lambda_1 + \lambda_2)  + 
       \c_{2\beta}^2 \left( \c^2_{\xi} \lambda_5 + \s^2_{\xi} \lambda_6 
                     \right)       
&
      \s_{2\beta}
      \left[ -2\c^2_{\beta} \lambda_1 + 2\s^2_{\beta} \lambda_2 + 
       \c_{2\beta} \left(\c^2_{\xi}\lambda_5 + \s^2_{\xi}\lambda_6
          \right)
       \right]
&    \frac12
      \c_{2\beta}\s_{2\xi} 
      \left( \lambda_6 - \lambda_5 \right)
\\
      \s_{2\beta}
      \left[ -2\c^2_{\beta} \lambda_1 + 2\s^2_{\beta} \lambda_2 + 
       \c_{2\beta} \left(\c^2_{\xi}\lambda_5 + \s^2_{\xi}\lambda_6
          \right) 
      \right]
& 
     4 \left[\c^4_{\beta}\lambda_1 + \s^4_{\beta}\lambda_2 + 
             \lambda_3 + \c^2_{\beta}\s^2_{\beta} 
             \left( \c^2_{\xi}\lambda_5 + \s^2_{\xi}\lambda_6 
             \right)  
       \right] 
&
  \frac12
    \s_{2\beta} \s_{2\xi}
    \left( \lambda_6 - \lambda_5 \right) 
\\
   \frac12 \c_{2\beta} \s_{2\xi}
                \left( \lambda_6 - \lambda_5 \right)          
&
  \frac12     
     \s_{2\beta} \s_{2\xi}
    \left( \lambda_5 - \lambda_6 \right)      
&  
    \s^2_{\xi}\lambda_5 + \c^2_{\xi}\lambda_6    
  \end{smallmatrix}
  \right)
  \label{eq:matrixm0}
\end{equation}
 By diagonalizing it, one gets the true Higgs bosons $h, H,A^0$. The operation
of charge conjugation $C$ means the complex conjugation of these physical fields
(i.e. not of those appearing in the parametrization \eqref{eq:param}).
However, we can employ the representation \eqref{eq:inaparam} involving fields
that are linear combinations of real variables without complex coefficients.
Note that for $\xi=0$ (the \CP-conserving case) the $A$ is a \CP-odd Higgs boson 
($A'$ = $A$ in such a case) and $H$, $h$ are \CP\ even. Such a statement is also true when
$\xi=\pi/2$ and/or $\lambda_5=\lambda_6$; as we shall see later in this
section, for these particular values of parameters there is again no \CP\
violation in the potential \eqref{eq:potential}. 
 
 For $\xi=0$ the Higgs boson masses can be calculated explicitly, and
subsequently one can express the coupling constants $\lambda_i$
in terms of masses and a mixing angle defined through
\begin{equation}
\doublet{h_1}{h_2} = \begin{pmatrix} \cos\alpha & -\sin\alpha \\
                                      \sin\alpha & \phantom{-}\cos\alpha \end{pmatrix}
                      \doublet hH \\
\label{eq:defmhmH}
\end{equation}
Let us now express the $\lambda_{1,2,3,4}$ in terms of the Higgs boson masses
in the case $\xi=0$
(as we have only four distinct masses, we leave the $\lambda_5$ as a
free parameter). One gets
\begin{equation}
 \begin{aligned}
  \lambda_4&= 2 v^{-2} m_\pm^2 \qquad 
  \lambda_6 = 2 v^{-2} m_A^2 \qquad
  \lambda_3 = 2 v^{-2} \frac{s_\alpha c_\beta}{s_\beta c_\beta} (m_H^2 -  m_h^2) -
              \frac{\lambda_5}4 \\
  \lambda_1&= \frac12 v^{-2} \left[ c_\alpha^2 m_H^2 + s_\alpha^2  m_h^2  	   
  				    - \frac{s_\alpha c_\beta}{\tan\beta} (m_H^2 - m_h^2) \right]
             -\frac{\lambda_5}4 \left( \frac1{\tan^2 \beta} - 1 \right) \\
  \lambda_2&= \frac12 v^{-2} \left[ s_\alpha^2 m_H^2 + c_\alpha^2  m_h^2  	    
  				    - s_\alpha c_\beta{\tan\beta} (m_H^2 - m_h^2) \right]
             -\frac{\lambda_5}4 \left( \frac1{\tan^2 \beta} - 1 \right) 
 \end{aligned}
 \label{eq:lambdatomasses}
\end{equation}
 Note also that the matrix of the quadratic form of the
 scalar fields is the Hessian of the potential at its minimum.
 The condition for the existence of a minimum is that the Hessian
 is positive definite, and this in turn means that
 the Higgs boson masses (squared) are positive.

   Finally, let us discuss briefly the particular cases $\xi=0$, $\xi=\pi/2$
and $\lambda_5=\lambda_6$. The case $\xi=0$ represents a
model without \CP\ violation within the scalar sector, as it is described in
\cite{Guide}. The case $\xi=\pi/2$ can be analyzed easily in the
parametrization \eqref{eq:inaparam}; using this, the potential can be viewed as the case
$\xi=0$ with the change of notation 
 \begin{equation}
  \Phi'_1 = \Phi_1 \qquad \Phi'_2 = \imag \Phi_2 \qquad 
  \lambda_5\leftrightarrow\lambda_6
 \end{equation}
 Thus, the two cases are equivalent. When $\lambda_6=\lambda_5$, the $\xi$-dependent
 part of the potential  can be recast as
 \begin{equation}
    \lambda_5 \left( \Re(\Phi_1^\dg \Phi_2) - \frac{v_1 v_2}{2} \cos{\xi} \right)^2
   +\lambda_6 \left( \Im(\Phi_1^\dg \Phi_2)  - \frac{v_1 v_2}{2} \sin{\xi} \right)^2 =
    \lambda_6 \left\rvert \Phi_1^\dg \Phi_2 - \frac{v_1 v_2}{2} \e^{\imag \xi} \right\lvert^2
 \end{equation}
 The remaining terms do not depend on the relative phase between $\Phi_1$ and
 $\Phi_2$, so that the phase factor $\e^{\imag\xi}$ can be transformed away and one
 thus again has a \CP-conserving case. A particular consequence of such an
 analysis is that for $\nu=0$ there can be no \CP\ violation. 
     
\section{LQT method}
\label{sec:LQT}
 For finding the upper bounds on the Higgs boson masses we will employ 
 the well-known LQT method invented three decades
 ago \cite{LQT}. This method relies on imposing the condition of 
 perturbative (in particular, tree-level) unitarity on an appropriate 
 set of physical scattering processes. Within a renormalizable theory, 
 the scattering amplitudes are "asymptotically flat", i.e. they do not exhibit
 any power-like growth in the high-energy limit. However, the dominant couplings
 are typically proportional to the scalar boson masses and one can thus obtain
 useful technical constraints on their values. In the pioneering paper 
 \cite{LQT} the method was applied to the minimal SM, and several groups of 
 authors employed it subsequently within models involving an extended Higgs 
 sector, in particular the THDM (cf.~\cite{OldPapers},
\cite{KKT}, \cite{AAN}). The results
 of various authors differ slightly, so it perhaps makes sense to reconsider 
 the corresponding calculation and present, for the sake of clarity, some
 additional technical details of the whole procedure. 

 In the spirit of the LQT approach, our analysis is based on the condition
 of tree-level $S$-matrix unitarity within the subspace of two-particle states.
 Instead of the unitarity condition used in the original paper \cite{LQT}, 
 we can adopt an improved constraint for the $s$-wave partial amplitude $\M_0$, 
 namely 
 \begin{equation}
  \left|\Re\M_0\right| \le \frac12
  \label{eq:podunitarity}
 \end{equation}
 (cf. \cite{Marciano:1989ns}). Note that the 
 tree-level matrix elements in question
 are real, and in the high-energy limit their leading contributions 
 do not involve any angular dependence. Thus, the $\M_0$ 
 generally coincides with the full tree-level (asymptotic) matrix element $\M$,
 up to a~conventional normalization factor of $16\pi$ appearing in 
 the standard partial-wave expansion. The effective unitarity 
 constraint \eqref{eq:podunitarity} then becomes 
\begin{equation}
 |\M| \le 8\pi
 \label{eq:osempi}
\end{equation}
 For an optimal implementation of the unitarity constraints we will consider the
eigenvalues of the matrix $M_{ij}=\M_{i\to j}$ where the indices $i$ and $j$ 
label symbolically all possible two-particle states. Having in mind our primary
goal, we take into account only binary processes whose matrix elements involve
the Higgs boson masses in the leading order, in particular in the $O(E^0)$
terms.  Invoking arguments analogous to those used in the original paper \cite{LQT},
one can
show that the relevant contributions descend from the interactions of Higgs
scalars and longitudinal vector bosons. Using the equivalence theorem for
longitudinal vector bosons and Goldstone bosons (see~e.g.~\cite{LQT}, 
\cite{Equivalence}) one finds out, in accordance with the LQT
treatment, that the only relevant contributions come from the amplitudes
involving Higgs bosons and unphysical Goldstone bosons (that occur in an
$R$-gauge formulation of the theory). It means that we will examine the
above-mentioned matrix $M_{ij}$ , including all two-particle states made of the
scalars (both physical and unphysical) $w^\pm, z, H^\pm, A^0, H, h$. It is not
difficult to see that the leading terms in the individual amplitudes are
determined by the direct (contact) quartic scalar interactions, while the
triple vertices enter second order Feynman graphs and their contributions are
suppressed by the propagator effects in the  high energy expansion.

   As noted above, we will be mainly concerned with the eigenvalues of the
two-particle scattering matrix. It means that for our purpose we can consider,
equivalently, any unitary transformation of the matrix $M_{ij}$.
 In particular, it
is more convenient to take, instead of the $M_{ij}$, a matrix consisting of the
scattering amplitudes between the two-particle states made of the "particles"
$w^\pm_a, z_a, h_a$ corresponding to the parametrization \eqref{eq:param}.
The eigenvalues of this matrix can be found in the earlier paper \cite{AAN}.

 Matrix elements for the scattering processes corresponding to the 
two-particle states
$(w_1^+ w_2^-, w_2^+ w_1^-,$ $ h_1 z_2, h_2 z_1,
z_1 z_2, h_1 h_2)$ form the submatrix
\begin{equation}
  \bordermatrix{
  &\m w_1^+ w_2^-&\m  w_2^+ w_1^-&\m h_1 z_2&\m h_2 z_1&\m z_1 z_2&\m h_1 h_2 \cr\vbox{\hrule}
\m w_1^+ w_2^- & 2 {{\lambda }_3} + \frac{{{\lambda }_5}}{2} + 
   \frac{{{\lambda }_6}}{2} & 4 
   \left( \frac{{{\lambda }_5}}{4} - 
     \frac{{{\lambda }_6}}{4} \right)  & \frac{i }
     {2} {{\lambda }_4} - 
   \frac{i }{2} {{\lambda }_6} & \frac{-i }{2} 
    {{\lambda }_4} + \frac{i }{2} {{\lambda }_6} &
   \frac{-{{\lambda }_4}}{2} + 
   \frac{{{\lambda }_5}}{2} & \frac{-{{\lambda }_4}}
    {2} + \frac{{{\lambda }_5}}{2}                     
   \cr
\m w_2^+ w_1^-& 4 \left( \frac{{{\lambda }_5}}{4} - 
     \frac{{{\lambda }_6}}{4} \right)  & 2 
    {{\lambda }_3} + \frac{{{\lambda }_5}}{2} + 
   \frac{{{\lambda }_6}}{2} & \frac{-i }{2} 
    {{\lambda }_4} + \frac{i }{2} {{\lambda }_6} &
   \frac{i }{2} {{\lambda }_4} - 
   \frac{i }{2} {{\lambda }_6} & \frac{-{{\lambda }_
        4}}{2} + \frac{{{\lambda }_5}}{2} & \frac{-{
          {\lambda }_4}}{2} + \frac{{{\lambda }_5}}{2}
   \cr
\m h_1 z_2 & \frac{-i }{2} {{\lambda }_4} + 
   \frac{i }{2} {{\lambda }_6} & \frac{i }{2} 
    {{\lambda }_4} - \frac{i }{2} {{\lambda }_6} &
   4 \left( \frac{{{\lambda }_3}}{2} + 
     \frac{{{\lambda }_6}}{4} \right)  & \frac{{{\lambda
         }_5}}{2} - \frac{{{\lambda }_6}}{2} & 0 & 0 
   \cr
\m h_2 z_1 & \frac{i }{2} {{\lambda }_4} - 
   \frac{i }{2} {{\lambda }_6} & \frac{-i }{2} 
    {{\lambda }_4} + \frac{i }{2} {{\lambda }_6} &
   \frac{{{\lambda }_5}}{2} - 
   \frac{{{\lambda }_6}}{2} & 4 
   \left( \frac{{{\lambda }_3}}{2} + 
     \frac{{{\lambda }_6}}{4} \right)  & 0 & 0 
   \cr 
\m z_1 z_2 & \frac{-{{\lambda }_4}}{2} + \frac{{{\lambda }_5}}{2}
   & \frac{-{{\lambda }_4}}{2} + 
   \frac{{{\lambda }_5}}{2} & 0 & 0 & 4 
   \left( \frac{{{\lambda }_3}}{2} + 
     \frac{{{\lambda }_5}}{4} \right)  & \frac{{{\lambda
         }_5}}{2} - \frac{{{\lambda }_6}}{2} 
   \cr
\m h_1 h_2 &
     \frac{-{{\lambda }_4}}{2} + \frac{{{\lambda }_5}}{2}
   & \frac{-{{\lambda }_4}}{2} + 
   \frac{{{\lambda }_5}}{2} & 0 & 0 & \frac{{{\lambda }_
       5}}{2} - \frac{{{\lambda }_6}}{2} & 4 
   \left( \frac{{{\lambda }_3}}{2} + 
     \frac{{{\lambda }_5}}{4} \right)  \cr
   }
\end{equation}
with eigenvalues
\begin{equation}
  \begin{aligned}
   e_1 &= 2 \lambda_3 - \lambda_4 - \frac12 \lambda_5 + \frac52 \lambda_6 \\
   e_2 &= 2 \lambda_3 + \lambda_4 - \frac12 \lambda_5 + \frac12 \lambda_6 \\
   f_+ &= 2 \lambda_3 - \lambda_4 + \frac52 \lambda_5 - \frac12 \lambda_6 \\
   f_- &= 2 \lambda_3 + \lambda_4 + \frac12 \lambda_5 - \frac12 \lambda_6 \\
   f_1 &= f_2 = 2 \lambda_3 + \frac 12 \lambda_5 + \frac12 \lambda_6
  \end{aligned}
\end{equation}
Another submatrix is defined by means of the states
$(w_1^+ w_1^-, w_2^+ w_2^-,
\frac{z_1 z_1}{\sqrt 2}, \frac{z_2 z_2}{\sqrt 2}, \frac{h_1 h_1}{\sqrt
2}, \frac{h_2 h_2}{\sqrt 2})$; it reads
\begin{equation}
  \m\bordermatrix{
  &\m w_1^+ w_1^-&\m w_2^+ w_2^-&\m
  \frac{z_1 z_1}{\sqrt 2}&\m \frac{z_2 z_2}{\sqrt 2}&\m 
  \frac{h_1 h_1}{\sqrt2}&\m \frac{h_2 h_2}{\sqrt 2}  \cr\m
w_1^+ w_1^-&\m 4\left( {{\lambda }_1} + {{\lambda }_3}
     \right)  &\m 2{{\lambda }_3} + 
   \frac{{{\lambda }_5}}{2} + \frac{{{\lambda }_6}}{2} &\m 
   {\sqrt{2}}\left( {{\lambda }_1} + 
     {{\lambda }_3} \right)  &\m {\sqrt{2}}
   \left( {{\lambda }_1} + {{\lambda }_3} \right)  &\m 
   {\sqrt{2}}\left( {{\lambda }_3} + 
     \frac{{{\lambda }_4}}{2} \right)  &\m {\sqrt{2}}
   \left( {{\lambda }_3} + \frac{{{\lambda }_4}}{2}
     \right)  
   \cr\m
w_2^+ w_2^-&\m 2{{\lambda }_3} + 
   \frac{{{\lambda }_5}}{2} + \frac{{{\lambda }_6}}{2} &\m 
   4\left( {{\lambda }_2} + {{\lambda }_3} \right)
      &\m {\sqrt{2}}\left( {{\lambda }_3} + 
     \frac{{{\lambda }_4}}{2} \right)  &\m {\sqrt{2}}
   \left( {{\lambda }_3} + \frac{{{\lambda }_4}}{2}
     \right)  &\m {\sqrt{2}}
   \left( {{\lambda }_2} + {{\lambda }_3} \right)  &\m 
   {\sqrt{2}}\left( {{\lambda }_2} + 
     {{\lambda }_3} \right)  
   \cr
\frac{z_1 z_1}{\sqrt 2} &\m 
   {\sqrt{2}}
   \left( {{\lambda }_1} + {{\lambda }_3} \right)  &\m 
   {\sqrt{2}}\left( {{\lambda }_3} + 
     \frac{{{\lambda }_4}}{2} \right)  &\m 3
   \left( \lambda_1 + 
     \lambda_3\right)  &\m 
    \lambda_1 + \lambda_3  &\m 
   \lambda_3 + \frac{{{\lambda }_5}}{2}   &\m 2
   \left( \frac{{{\lambda }_3}}{2} + 
     \frac{{{\lambda }_6}}{4} \right)  
   \cr
\frac{z_2 z_2}{\sqrt 2}&\m   {\sqrt{2}}
   \left( {{\lambda }_1} + {{\lambda }_3} \right)  &\m 
   {\sqrt{2}}\left( {{\lambda }_3} + 
     \frac{{{\lambda }_4}}{2} \right)  &\m 2
   \left( \frac{{{\lambda }_1}}{2} + 
     \frac{{{\lambda }_3}}{2} \right)  &\m 3
   \left(\lambda_1 + \lambda_3 \right)  &\m 2
   \left( \frac{{{\lambda }_3}}{2} + 
     \frac{{{\lambda }_6}}{4} \right)  &\m 2
   \left( \frac{{{\lambda }_3}}{2} + 
     \frac{{{\lambda }_5}}{4} \right)  
   \cr
\frac{h_1 h_1}{\sqrt2}&\m 
   {\sqrt{2}}
   \left( {{\lambda }_3} + \frac{{{\lambda }_4}}{2}
     \right)  &\m {\sqrt{2}}
   \left( {{\lambda }_2} + {{\lambda }_3} \right)  &\m 
   2\left( \frac{{{\lambda }_3}}{2} + 
     \frac{{{\lambda }_5}}{4} \right)  &\m 2
   \left( \frac{{{\lambda }_3}}{2} + 
     \frac{{{\lambda }_6}}{4} \right)  &\m 3
   \left( \lambda_2 + \lambda_3 \right)  &\m 2
   \left( \frac{{{\lambda }_2}}{2} + 
     \frac{{{\lambda }_3}}{2} \right)  
   \cr 
\frac{h_2 h_2}{\sqrt 2}&\m    {\sqrt{2}}
   \left( {{\lambda }_3} + \frac{{{\lambda }_4}}{2}
     \right)  &\m {\sqrt{2}}
   \left( {{\lambda }_2} + {{\lambda }_3} \right)  &\m 
   2\left( \frac{{{\lambda }_3}}{2} + 
     \frac{{{\lambda }_6}}{4} \right)  &\m 2
   \left( \frac{{{\lambda }_3}}{2} + 
     \frac{{{\lambda }_5}}{4} \right)  &\m 2
   \left( \frac{{{\lambda }_2}}{2} + 
     \frac{{{\lambda }_3}}{2} \right)  &\m 3
   \left( \lambda_2 + \lambda_3 \right)  \cr  
  }
\end{equation}
and its eigenvalues are
\begin{equation}
  \begin{aligned}
    a_\pm & = 3 (\lambda_1 + \lambda_2 + 2\lambda_3) \pm
    \sqrt{9(\lambda_1 - \lambda_2)^2 + 
      [4\lambda_3 + \lambda_4 + \tfrac12 (\lambda_5 + \lambda_5)]^2} \\
    b_\pm & = \lambda_1 + \lambda_2 + 2\lambda_3 \pm 
    \sqrt{(\lambda_1-\lambda_2)^2 + 
    \tfrac14 (-2 \lambda_4 + \lambda_5 + \lambda_6)^2} \\
    c_\pm & = \lambda_1 + \lambda_2 + 2 \lambda_3 \pm 
    \sqrt{(\lambda_1 - \lambda_2)^2 + \tfrac 14 (\lambda_5 - \lambda_6)^2} \\
  \end{aligned}
  \label{eq:defabc}
\end{equation}
A third submatrix
 \begin{equation}
   \bordermatrix{
    &\m h_1 z_1&\m h_2 z_2 \cr
\m h_1 z_1&2\left( \lambda_2 + 
     \lambda_3 \right)  & \tfrac12 (\lambda_5 - \lambda_6) \cr 
\m h_2 z_2 &  \tfrac12(\lambda_5 - \lambda_6)
   & 2\left( \lambda_1 + \lambda_3 \right)  \cr
   }
 \end{equation}
has eigenvalues $c_\pm$ (see \eqref{eq:defabc}).
Finally, there are submatrices corresponding to 
charged states
$(h_1 w_1^+$, $h_2 w_1^+$, $z_1 w_1^+$, $z_2 w_1^+$,
  $h_1 w_2^+$, $h_2 w_2^+$, $z_1 w_2^+$, $z_2 w_2^+)$:
 \begin{equation}
   \bordermatrix {
    &\m h_1 w_1^+&\m h_2 w_1^+&\m z_1 w_1^+&\m z_2 w_1^+ \cr
\m h_1 w_1^+ &    2\left( {{\lambda }_1} + {{\lambda }_3}
     \right)  & \frac{-{{\lambda }_4}}{2} + 
   \frac{{{\lambda }_5}}{2} & 0 & \frac{i }{2}
    {{\lambda }_4} - \frac{i }{2}{{\lambda }_6} 
   \cr
\m h_2 w_1^+ &
   \frac{-{{\lambda }_4}}{2} + 
   \frac{{{\lambda }_5}}{2} & 2
   \left( {{\lambda }_2} + {{\lambda }_3} \right)  &
   \frac{i }{2}{{\lambda }_4} - 
   \frac{i }{2}{{\lambda }_6} & 0 
   \cr
\m z_1 w_1^+ &   0 & 
    \frac{-i }{2}{{\lambda }_4} + 
   \frac{i }{2}{{\lambda }_6} & 2
   \left( {{\lambda }_1} + {{\lambda }_3} \right)  &
   \frac{-{{\lambda }_4}}{2} + 
   \frac{{{\lambda }_5}}{2} \cr
\m z_2 w_1^+ & \frac{-i }{2}
    {{\lambda }_4} + \frac{i }{2}{{\lambda }_6} & 
   0 & \frac{-{{\lambda }_4}}{2} + 
   \frac{{{\lambda }_5}}{2} & 2
   \left( {{\lambda }_2} + {{\lambda }_3} \right)  \cr
   }
 \end{equation}
\begin{equation}
 \bordermatrix{
   &\m h_1 w_2^+&\m  h_2 w_2^+&\m  z_1 w_2^+&\m  z_2 w_2^+ \cr
\m h_1 w_2^+&2\left( {{\lambda }_3} + 
    \frac{{{\lambda }_4}}{2} \right)  & \frac{-{{\lambda
          }_4}}{2} + \frac{{{\lambda }_5}}{2} & 0 & 
   \frac{i }{2}{{\lambda }_4} - 
   \frac{i }{2}{{\lambda }_6} 
   \cr
\m h_2 w_2^+&   \frac{-{{\lambda
          }_4}}{2} + \frac{{{\lambda }_5}}{2} & 2
   \left( {{\lambda }_3} + \frac{{{\lambda }_4}}{2}
     \right)  & \frac{i }{2}{{\lambda }_4} - 
   \frac{i }{2}{{\lambda }_6} & 0
   \cr
\m z_1 w_2^+& 0 & 
    \frac{-i }{2}{{\lambda }_4} + 
   \frac{i }{2}{{\lambda }_6} & 2
   \left( {{\lambda }_3} + \frac{{{\lambda }_4}}{2}
     \right)  & \frac{-{{\lambda }_4}}{2} + 
   \frac{{{\lambda }_5}}{2} 
   \cr 
\m z_2 w_2^+ &   \frac{-i }{2}
    {{\lambda }_4} + \frac{i }{2}{{\lambda }_6} & 
   0 & \frac{-{{\lambda }_4}}{2} + 
   \frac{{{\lambda }_5}}{2} & 2
   \left( {{\lambda }_3} + \frac{{{\lambda }_4}}{2}
     \right)  \\    
 }
\end{equation}
Their eigenvalues are the
$f_-$, $e_2$, $f_1$, $c_\pm$, $b_\pm$ shown above and,
in addition, 
\begin{equation}
  p_1 = 2 (\lambda_3 + \lambda_4) - \frac 12 \lambda_5 - \frac 12 \lambda_6
\end{equation}
 Unitarity conditions \eqref{eq:osempi} for the eigenvalues
 listed above give the constraints
\begin{equation}
 |a_\pm|, |b_\pm|, |c_\pm|, |f_\pm|, |e_{1,2}|, |f_1|, |p_1| \le 8\pi
 \label{eq:inequalities}
\end{equation}
 Note that an independent derivation of these inequalities 
 based on symmetries of the Higgs potential can be found in the papers
 \cite{Ginzburg:2003,Ginzburg:2004}.

\section{Independent inequalities}
\label{sec:inequalities}

However, the inequalities \eqref{eq:inequalities} are not all independent.
Indeed, it is not difficult to observe some simple relations as 
\begin{equation}
 \begin{aligned}
 3 f_1 &=   p_1 + e_1 + f_+ \\
 3 e_2 &= 2 p_1 + e_1 \\
 3 f_- &= 2 p_1 + f_+
 \end{aligned}
 \label{vztahyfef}
\end{equation}
and this means that the inequalities $|p_1|, |f_+|, |e_1| \le 8\pi$ imply
$|f_1|, |e_2|, |f_-|\le 8\pi$. Further, the eigenvalues \eqref{eq:defabc}
 in the remaining inequalities can be rewritten as
\begin{equation}
 \begin{aligned}
 a_\pm &= 3 \lambda_{123} \pm \sqrt{(3\lambda_{12})^2 + \tfrac14(f_++e_1+2p_1)^2} \\
 b_\pm &=  \lambda_{123} \pm \sqrt{(\lambda_{12})^2 +\tfrac1{36}(f_++e_1-2p_1)^2} \\
 c_\pm &=  \lambda_{123} \pm \sqrt{(\lambda_{12})^2 + \tfrac1{36}(f_+ - e_1)^2} 
 \end{aligned}
 \label{eq:newabc}
\end{equation}
 where $\lambda_{123}=\lambda_1 + \lambda_2 + 2 \lambda_3$ and
$\lambda_{12}=\lambda_1-\lambda_2$. 
In the case $\lambda_{123}>0$ the inequalities for the $a_-$, $b_-$, $c_-$
follow from $a_+, b_+, c_+ \le 8\pi$. For $\lambda_{123}<0$ the situation
is similar,  with interchanges $(a,b,c)_\pm\to (a,b,c)_\mp$ and
$\lambda_{123}\to -\lambda_{123}$.

The authors \cite{KKT} noticed that among the latter inequalities, the strongest
one is $a_+<8\pi$; Indeed, using $\eqref{eq:inequalities}$ and \eqref{eq:newabc} 
one can show that for  $\lambda_{123}>0$ the remaining ones follow from it. In the
case $\lambda_{123}<0$ the same statement is true concerning $a_+<8\pi$.

   Thus, it is sufficient to solve the inequalities
\begin{equation}
 |a_\pm|, |f_+|, |e_1|, |p_1| \le 8\pi
 \label{uniq}
\end{equation}

In fact, the inequality $a_{-}<8\pi$ need not be taken into account in subsequent discussion;
it turns out that this is weaker than the remaining ones and does not influence bounds in 
question (one can verify \textit{a posteriori} that our solutions satisfy the constraints
$a_{-}<8\pi$ automatically).

\section[Upper bounds for MA and M+- with xi=0]{Upper bounds for $M_A$ and $M_\pm$ with $\xi=0$}
\label{sec:MaMpm}

Before starting our calculation, let us recall that the condition $\xi = 0$
means that the $Z_2$ symmetry-breaking parameter $\nu$ becomes $\nu =
\lambda_5$ (see \eqref{eq:nu}). To
proceed, we shall first fix convenient notations. The LQT bound for the
SM Higgs mass sets a natural scale for our estimates, so let us introduce it
explicitly: 
\begin{equation}
 m_\text{LQT}=\sqrt{\frac{4\pi\sqrt2}{3G_\text{F}}}=
   \sqrt{\frac{8\pi}{3}} v \doteq 712 \GeV
 \label{eq:mlqt}
\end{equation}
(note that in writing eq.\eqref{eq:mlqt} we do not stick strictly to the original
value \cite{LQT}, using rather the improved bound \cite{Marciano:1989ns}). 
In the subsequent
discussion we shall then work with the dimensionless ratios 
\begin{equation}
 M = \frac{m}{m_\text{LQT}}
\end{equation}
instead of the true scalar boson masses (denoted here generically as $m$).
Further, an overall constant factor $16\pi/3$ can be absorbed in a convenient
redefinition of the coupling constants, by writing
\begin{equation}
 \lambda'_i=\frac{3\lambda_i}{16\pi}
 \label{lambdaprime}
\end{equation}
Finally, we introduce new variables
\begin{equation}
 X=M^2_H+M^2_h, \quad Y=M^2_H - M^2_h, \quad Z=\frac{\sin 2\alpha}{\sin 2\beta} Y
\end{equation}
that will help to streamline a bit the solution of the inequalities in question.

Using equations \eqref{eq:lambdatomasses} and the definitions shown above, the
$\lambda'$ can be expressed as
\begin{equation}
  \begin{aligned}
        \lambda'_4 &= M^2_\pm \\
        \lambda'_6 &= M^2_A \\
        \lambda'_3 &= 
            \frac14 \frac{\sin 2\alpha}{\sin 2\beta} Y 
            - \frac14\lambda'_5 = \frac Z4 - \frac{\lambda'_5}4 \\
        \lambda'_{12} &= \frac1{2\sin^2{2\beta}} 
       \left[ (X-2\lambda'_5) \cos 2 \beta - Y \cos 2\alpha\right] \\
        \lambda'_{123} &= \frac1{2\sin^22\beta} 
         (X - Y \cos 2\alpha \cos  2\beta - 2\lambda'_5)+
	 \frac{\lambda'_5}2 
  \end{aligned}
  \label{eq:lambdaparam}
\end{equation}
Let us now discuss the possible bounds for the $M_\pm, M_A$.
These can be obtained from the inequalities for $|e_1|,|f_+|, |p_1|$, which
read, in our new notation
\begin{equation}
\begin{aligned}
   \left| \frac Z2 - \lambda'_5 - M^2_\pm + \frac52 M^2_A \right| 
     &\le \frac32 \\
   \left| \frac Z2 + 2 \lambda'_5 - M^2_\pm - \frac12 M^2_A \right| 
     &\le \frac32 \\   
   \left| \frac Z2 -  \lambda'_5 + 2M^2_\pm - \frac12 M^2_A \right| 
     &\le \frac32 \\
\end{aligned}     
 \label{eq:mxima1}
\end{equation}
The relations \eqref{eq:mxima1} are linear with respect to the 
$M^2_\pm, M^2_A$ and one
can thus view the domain defined by these inequalities as a hexagon in the
plane $(M^2_\pm,M^2_A)$. Then it is clear that the highest possible value of a
mass variable in question will correspond to a vertex (or a~whole hexagon
side). By examining all possible cases one finds easily
that for $M^2_\pm$, such a~"critical" vertex satisfies the condition
$-f_+=p_1=8\pi$; in view of \eqref{eq:mxima1} this means that it corresponds to the
values
\begin{equation}
 (M^2_\pm, M^2_A) = (1+\lambda'_5, 1 + Z + 2 \lambda'_5)
 \label{eq:mxima2}
\end{equation}
Such a maximum value of  the $M^2_\pm$ is indeed formally admissible (in the
sense that by reaching it one does not leave the parametric space of the
considered model). To see this, one can substitute in eq. \eqref{eq:mxima2}
$M^2_A= \lambda'_5 , M^2_H=1 + \lambda'_5, M^2_h=0, \alpha=\pi-\beta$. Thus, the
bound becomes
\begin{equation}
 M^2_\pm \le 1 + \lambda'_5
 \label{Mxibound}
\end{equation}
Similarly, for $M^2_A$ the extremal solution corresponds to a hexagon vertex
defined by $e_1=-f_+=8\pi$ and its coordinates in the 
$(M^2_\pm, M^2_A)$ plane are then
\begin{equation}
 (M^2_\pm, M^2_A) = (1+\frac Z2+\frac32 \lambda'_5, 1  +  \lambda'_5)
\end{equation}
The parameter values that saturate this maximum are analogous and one has to take
$M^2_\pm= \lambda'_5/2 , M^2_H=1 + \lambda'_5, M^2_h=0, \alpha=\pi-\beta$.
In this way, the bound for $M^2_A$ becomes
the same as that for the $M^2_\pm$ , namely
\begin{equation}
 M^2_A \le 1 + \lambda'_5
 \label{MAbound}
\end{equation}

\section[Upper bounds for Mh, MH with xi=0]{Upper bounds for $M_h, M_H$ with $\xi=0$}
\label{sec:MhMH}

Let us now proceed to discuss the upper bounds for $M_H$ and $M_h$.
 If we considered the relevant constraints without any
further specification of the scalar bosons $h$ and $H$, we would get the same
result for both particles, since their interchange corresponds just to the
replacement $\alpha \to -\alpha$ (cf. eq. \eqref{eq:defmhmH}). Thus, let us add the
condition $M_h \le M_H$ (i.e.~$Y > 0$). In such a case, we will solve
just the inequality $a_+<8\pi$  (which puts the most stringent
bounds on the variables $X, Y$) and in the obtained solution we will constrain
the $M_A, M_\pm$ so as to satisfy the rest of the inequalities.

   The basic constraint $a_+<8\pi$ is quadratic with respect to
the $X, Y$ and reads (cf. the expression \eqref{eq:defabc})
\begin{multline}
  (X-Y \cos 2\alpha \cos 2\beta) - \lambda'_5(2-\sin^2 2\beta) + \\
    \sqrt{\big[ (X-2\lambda'_5)\cos2\beta - Y \cos 2\alpha  \big]^2
     + \left(\frac23\right)^2 \sin^4 2\beta \Big(
      Y\frac {\sin2\alpha}{\sin 2\beta} - \frac {\lambda'_5}2 
       + M^2_\pm + \frac {M^2_A}2
        \Big)^2} \le \sin^2 2\beta
  \label{eq:aplus}
\end{multline}
To work it out, we will employ the following trick: As a first step, we
will consider a simpler inequality, which is obtained from (34) by
discarding the second term under the square root; in other words, we will
first assume that
\begin{equation}
 Y\frac {\sin2\alpha}{\sin 2\beta} - \frac {\lambda'_5}2 
       + M^2_\pm + \frac {M^2_A}2 = 0
 \label{eq:apluszerocondition} 
\end{equation}
Of course, the "reduced" constraint
\begin{equation}
 X-Y \cos2\alpha \cos2\beta - \lambda'_5 (2-\sin^22\beta) +
 \left| X\cos2\beta - Y\cos2\alpha - 2\lambda'_5 \cos2\beta  \right| \le \sin^2 2\beta
 \label{eq:aplussimple}
\end{equation}
is in general weaker than the original one. Nevertheless, in a next step we
will be able to show that the obtained mass bound does get saturated for
appropriate values of the other parameters (such that the condition \eqref{eq:apluszerocondition} is
met) - i.e. that in this way we indeed get the desired minimum upper mass
bound corresponding to the original constraint \eqref{eq:aplus}. Thus, let us examine
the inequality \eqref{eq:aplussimple}. Obviously, we have to distinguish two possible cases:
\begin{enumerate}
 \item $ (X-2\lambda'_5)\cos2\beta \ge Y\cos2\alpha $. \par
  Then one has
  \begin{equation}
    X(1+\cos2\beta) - Y (1+\cos2\beta)\cos2\alpha - \lambda'_5 (1+\cos2\beta)
     \le \sin^22\beta
    \label{eq:aplusfirstcase}
  \end{equation}
  Making use of our assumption, we can get from \eqref{eq:aplusfirstcase} a simple constraint
  that does not involve $Y$, namely
  \begin{equation}
    X \le 1 + \lambda'_5
    \label{Xupperbound}
  \end{equation}
   (to arrive at the last relation, we had to divide by the factor
   $1-\cos2\beta$; when it vanishes, we can use directly the original
   inequality \eqref{eq:aplus} and get the same result).
 \item $(X-2\lambda'_5) \cos 2\beta \le Y \cos2\alpha$. \par 
   In a similar way as in the preceding case, the inequality
   \eqref{eq:aplussimple} implies the same bound
   \eqref{Xupperbound}.
\end{enumerate}
Thus, having constrained $X=M^2_H+M^2_h$ according to \eqref{Xupperbound}, we can
obviously also write
\begin{equation}
 M^2_H \le 1+\lambda'_5
 \label{MHbound}
\end{equation}
Now, it is not difficult to see that for
$M^2_h=0, M^2_\pm=\lambda'_5 + \tfrac12, 
M^2_A=1+\lambda'_5, \alpha=\pi-\beta$, eq.~\eqref{eq:apluszerocondition}
is satisfied with $M^2_H=1+\lambda'_5$ and means that
\eqref{MHbound} represents the mass upper bound pertinent to the
original unitarity constraint \eqref{eq:aplus}.

   The bound for the $M_h$ is obtained from \eqref{Xupperbound} by using there our subsidiary
condition $M_h \le M_H$; one thus has
\begin{equation}
 M^2_h \le \frac 12 (1+\lambda'_5)
 \label{Mhsmallbound}
\end{equation}
The upper limit in \eqref{Mhsmallbound} gets saturated (i.e.~$M^2_h=\frac 12
(1+\lambda'_5) $) for $M_H=M_h$, $M^2_A=0$, $M^2_\pm=\lambda'_5/2$, $\alpha=3\pi/4, \beta=\pi/4$.
It is worth noticing that
here we have fixed a particular value of the angle $\beta$ , while all previous
constraints were independent of $\beta$ (i.e. for any $\beta$ we were then able to find an
appropriate value of $\alpha$). A more detailed analysis shows that, in general,
the upper bound for the $M_h$ indeed depends explicitly on the $\beta$. To derive the corresponding
formula, we consider the boundary value $M_h = M_H$ (i.e.~$Y = 0$) and use also
eq.~\eqref{eq:apluszerocondition}. The inequality \eqref{eq:aplus} then
becomes \begin{equation}
 M^2_h - \lambda'_5\left(1-\frac{\sin^2 2\beta}2 \right) + 
 |M^2_h \cos2\beta -  \lambda'_5 \cos2\beta|
 \le \frac{\sin^22\beta}2
 \label{Mhbetadifficult}
\end{equation}
To work it out, we will assume that $M^2_h \ge \lambda'_5$ (taking into account
\eqref{Mhsmallbound} this means $\lambda'_5\le 1$; in fact, one can do even without such a
restriction, but for our perturbative treatment only sufficiently small
values of the $\lambda'_5$ are of real interest). The inequality \eqref{Mhbetadifficult} then
becomes
\begin{equation}
 M^2_h \le \frac{(1-\lambda'_5)}2
    \frac{(1+\cos 2\beta)(1-\cos2\beta)}{1+|\cos2\beta|}
 +\lambda'_5
 \label{Mhbeta}
\end{equation}
Obviously, the maximum bound \eqref{Mhsmallbound} is recovered from the last expression for
$\beta = \pi/4$. Let us also remark that the choice $\alpha = \pi-\beta$
comes, as in all previous cases, from the requirement $Z = - Y$.

\section[Upper bound for the lightest scalar for xi = 0]{Upper bound for the lightest scalar for $\xi = 0$}
\label{sec:Mlightest}

\begin{figure}[t]
 \centering
 \includegraphics{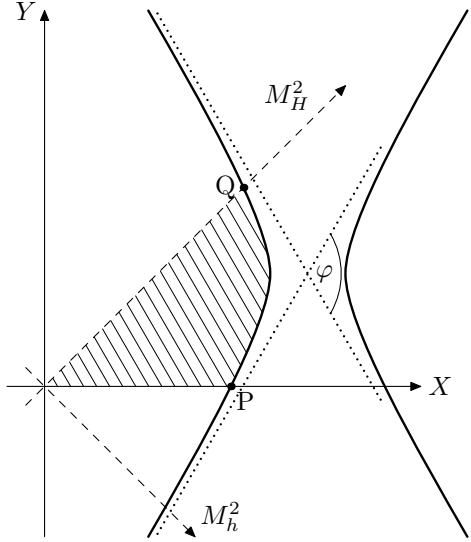}
 \caption{The region of admissible values of $M^2_h$ if $h$ is assumed to be the lightest scalar}
 \label{obrhyperbola}
\end{figure}

One can notice that any Higgs mass upper limit discussed so far gets saturated
only when at least one of the other scalar masses vanishes. Thus, another
meaningful question arising in this connection is what can be an upper bound
for the lightest Higgs boson (within a considered set of the five scalars $h, H,
A^0, H^\pm$). Let us first take $h$ to be the lightest scalar state; it means that in our
analysis we will include the additional assumption $M_h\le M_H, M_A, M_\pm$.
The procedure we are going to employ is a modest generalization of the earlier
calculation \cite{KKT}. Squaring the inequality \eqref{eq:aplus} one gets
\begin{equation}
 (X-X_0)^2 - \left(1 - \frac 59 \sin^2 2\alpha\right) (Y-Y_0)^2 \ge R^2
 \label{eq:hyperbola}
\end{equation}
where $X_0, Y_0$ and $R$ depend on $\lambda'_5, \alpha, \beta, M^2_\pm+M^2_A/2$. This
inequality defines the domain bounded by the hyperbola shown in Fig.~\ref{obrhyperbola}, but the
original constraint \eqref{eq:aplus} corresponds just to its left-hand part. In order to
find the solution, one should realize that the slope of the asymptote with
respect to the $X$-axis must be greater than the slope of the straight lines $X=\pm Y$
 (this follows from the fact that the coefficient
$1-\frac 59 \sin^2 2\alpha$, multiplying the $Y^2$ in \eqref{eq:hyperbola}, is less than one).
Because of that, the maximum value of the $M_h$ corresponds to $Y=0$ and
$a_+=8\pi$, and we are thus led to the equation
\begin{equation}
 X-\lambda'_5(2-\sin^22\beta) + \sqrt{\cos^22\beta(X-2\lambda'_5)^2 + 
  \frac 49 \sin^42\beta \left( M^2_\pm + \frac{M^2_A}2 - \frac{\lambda'_5}2\right) }
   = \sin^22\beta
 \label{Mhgeneral}
\end{equation}

It is clear that for smaller $M_\pm, M_A$ one has a bigger value of the $M_h$,
so the needed upper estimate is obtained for  $M_\pm = M_A = M_h$
(note also that from $Y=0$ one has $X = 2 M^2_h$). In this way one gets an
equation for maximum $M_h$:
\begin{equation}
 2 M^2_h - \lambda'_5(2-\sin^22\beta) + 
  \sqrt{ 4(M^2_h-\lambda'_5)^2\cos^22\beta + \sin^42\beta (M^2_h - \frac13\lambda'_5)^2} 
   = \sin^22\beta
   \label{Mhbetabound}
\end{equation}
From eq. \eqref{Mhbetabound} one can calculate the $M_h^2$ as a function of
$\sin^22\beta$.  It can be shown that for $\lambda'_5<3/5$ 
this function is increasing, i.e. the maximum is reached for $\beta=\pi/4$
and its value becomes

\begin{equation}
 M^2_h = \frac13 + \frac 49 \lambda'_5
 \label{Mhsmallestbound}
\end{equation}
We do not display the explicit dependence of the maximum $M_h$ on the $\beta$, but it is
clear that the solution of eq. $\eqref{Mhbetabound}$ is straightforward. Finally, we should also
examine the cases where the lightest Higgs boson mass is either $M_A$  or $M_\pm$  .
However, from the above discussion it is clear that both these extremes occur
when $M_h=M_A=M_\pm$.  
 \begin{figure}[t]
  \centering
  \includegraphics{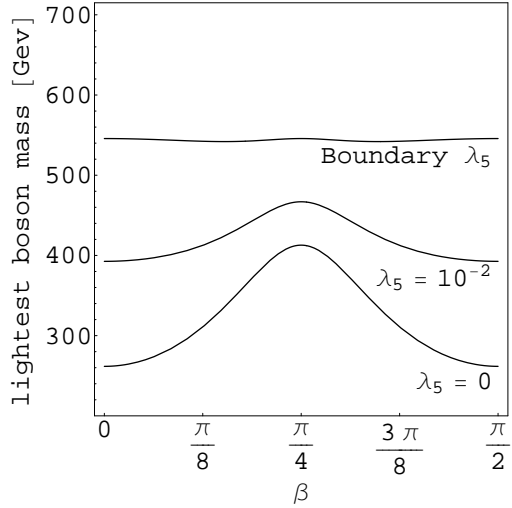}
  \caption{Dependence of lightest boson mass on $\beta$}
  \label{grafbeta}
 \end{figure}

Similarly, from eq. \eqref{Mhgeneral} one can derive a constraint
for the mass of the lightest neutral scalar boson (which we denote $M_n$). 
In this case we substitute there $X=2 M^2_n$, $M^2_A=M^2_n$, $M^2_\pm=0$ and obtain
thus the equation
\begin{equation}
 2 M^2_n - \lambda'_5(2-\sin^22\beta) + 
  \sqrt{ 4(M^2_n-\lambda'_5)^2\cos^22\beta + \sin^42\beta \left(\frac{M^2_n}3 - \frac13\lambda'_5\right)^2} 
   = \sin^22\beta
   \label{Mnbetabound}
\end{equation}
From eq. \eqref{Mnbetabound} one then obtains the $M_n^2$ as a function of
 $\sin^22\beta$, which is increasing for $\lambda'_5<1$. Its maximum reached at
 $\beta=\pi/4$ becomes
\begin{equation}
 M_n^2 = \frac 37 + \frac 47 \lambda'_5
 \label{Mnbound}
\end{equation}

 \section[Numerical solution for xi!=0 ]{Numerical solution for $\xi\neq0$}
\label{sec:numeric}

In the general case with $\xi\neq0$ (i.e. with \CP\ violation in the scalar
sector) we have not been able to solve the inequalities \eqref{uniq} analytically,
so we had to resort to an appropriate numerical procedure. The main result
we have obtained in this way is that for small values of the parameter $\nu$
(see eq. \eqref{eq:nu}), in particular for $\nu'\in \langle0,0.3\rangle$ 
 , the upper mass
bounds in question are the same as for $\xi = 0$. 
The interval has been chosen such that the variations in the upper estimates be
at the level of $50-100\%$, the validity of our theoretical estimates is guaranteed up
to $\nu'<3/5$ (see the remark below eq. \eqref{Mhbetabound}).

\begin{figure}[p]
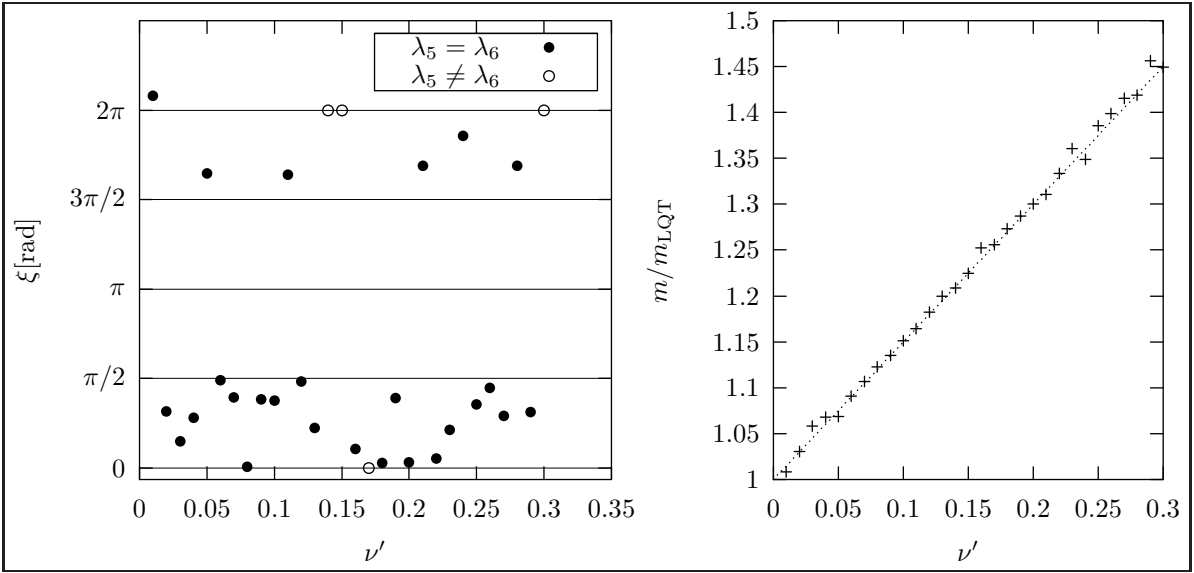

   \centering
   \fbox{\includegraphics{mxmasses.0}
   \includegraphics{mxmasses.1}}
  \caption{Charged Higgs boson, with theoretical estimate \eqref{Mxibound}}
  \label{fig:mx}
 \end{figure}

\begin{figure}[p]
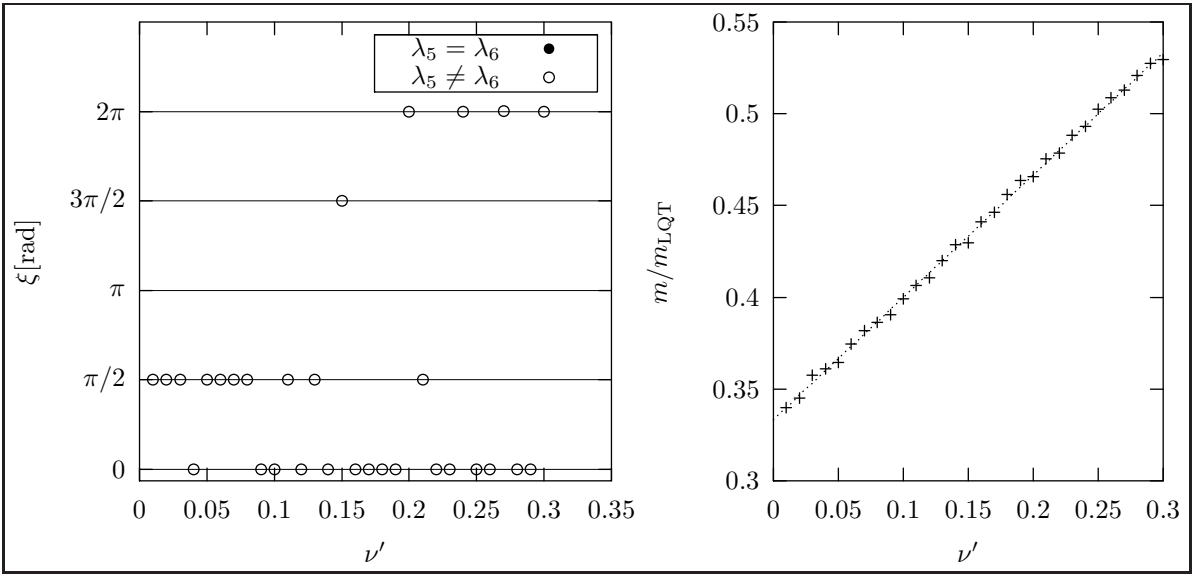

   \centering
   \fbox{\includegraphics{minmasses.0}
   \includegraphics{minmasses.1}}   
  \caption{Lightest scalar mass, with theoretical estimate \eqref{Mhsmallestbound}}
  \label{fig:min}
 \end{figure}

\begin{figure}[p]
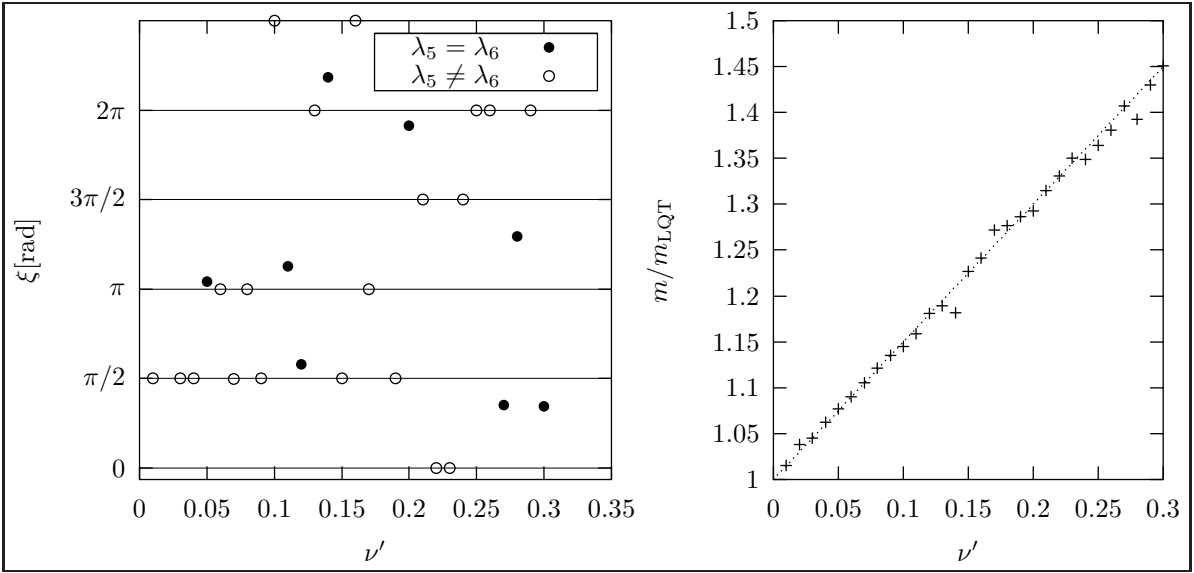

   \centering
   \fbox{\includegraphics{m1masses.0}
   \includegraphics{m1masses.1}}
  \caption{Heaviest neutral Higgs boson, with theoretical estimate
   \eqref{MHbound} or \eqref{MAbound}}
  \label{fig:m1}
 \end{figure}

\begin{figure}[p]
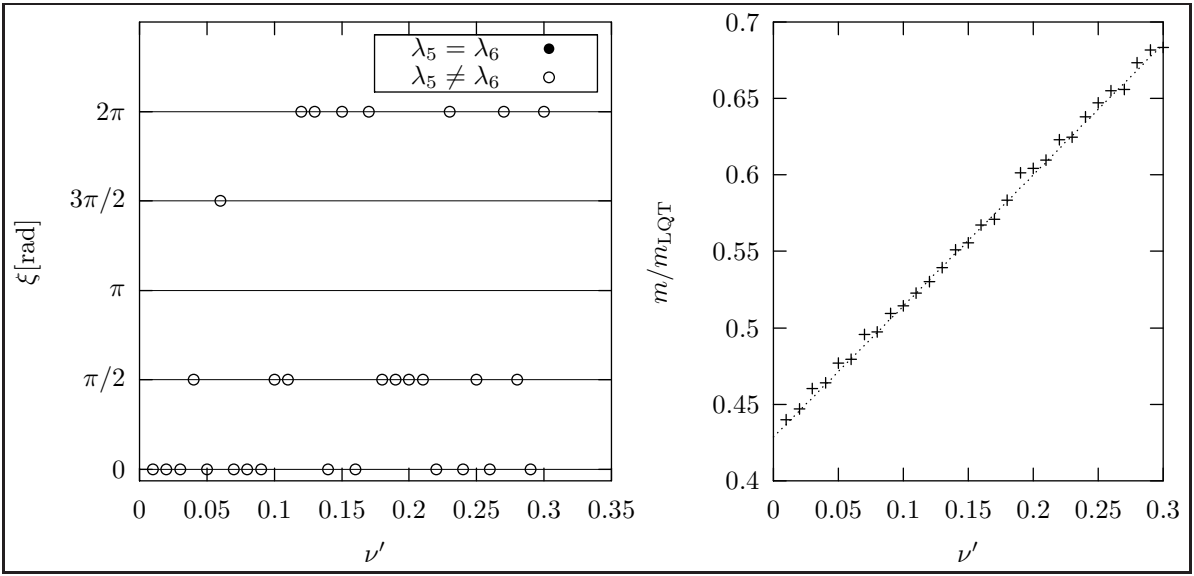

   \centering
   \fbox{\includegraphics{m3masses.0}
   \includegraphics{m3masses.1}}
   \caption{Lightest neutral boson with theoretical estimate \eqref{Mnbound}}
  \label{fig:m3}
 \end{figure}

   Our numerical procedure consists in solving the inequalities \eqref{uniq} on the
space of parameters $\lambda'_{1,2,3,4,5,6}$ and $\xi$ restricted by the
condition \eqref{eq:nu}, where one also adds constraints for the existence of
a minimum of the potential \eqref{eq:potential}: $\lambda'_4>0$ (i.e. $m^2_\pm>0$,
see \eqref{eq:lambdatomasses}) and the requirement of positive definiteness of the matrix 
\eqref{eq:matrixm0}
(i.e. $m^2_{A,H,h}>0$). On this parametric subspace we have looked for the
maximum values of the following quantities:

 \begin{enumerate}
   \item Mass of the charged Higgs boson $m_\pm$ (see Fig.~\ref{fig:mx})
   \item Mass of the lightest Higgs boson (see Fig.~\ref{fig:min})
   \item\label{item:lightest}
          Mass of the lightest neutral Higgs , i.e. the lightest one among the
         $A, H, h$ (see Fig.~\ref{fig:m1}) 
   \item Mass of the heaviest neutral Higgs, i.e. the heaviest among the $A, H, h$ (see Fig.~\ref{fig:m3}). 
 \end{enumerate}
 Let us remark that in this case we have not distinguished between $A$
 and $h, H$, which are superpositions of the \CP-odd and \CP-even states.

   In our plots we display, apart from the dependence of masses in question
on the $\nu$, also the values of the parameter $\xi$ in the case $\lambda_5=\lambda_6$ 
and $\lambda_5\neq\lambda_6$ respectively, in order to be able to 
distinguish the extreme cases without \CP\ violation ($\xi=k \pi/2$ or
$\lambda_5=\lambda_6$, see the discussion in Section \ref{sec:potential}). From Figs. \ref{fig:mx}, \ref{fig:min}, \ref{fig:m1},
 \ref{fig:m3} it
can be seen that all examined mass upper bounds are reached just in the
aforementioned extreme cases. In view of this, we can make use of our
previous analytic expressions, except for the case \ref{item:lightest}, which we have not
solved analytically.
 
   Our results have been simulated by means of the computer program Matlab
6.0, package optim, with the help of the function fmincon. The numerical
 errors are
mostly due to an insufficiently smooth condition for the positive
definiteness of the matrix \eqref{eq:matrixm0}.


\section{Conclusions}
\label{sec:conclusion}
  In the present paper we have reconsidered upper bounds for the scalar boson
masses within THDM, by using the well-known technical constraint of tree-level
 unitarity. Our analysis should extend and generalize the results of
some previous treatments, in particular those obtained in the papers \cite{AAN}
and \cite{KKT}. Although we basically employ the traditional methods, we have
tried to present some details of the calculations not shown in the earlier
papers --- we have done so not only for the reader's convenience, but also to provide a
better insight into the origin of the numerical results displayed here. 
As we have already noted in the Introduction, some new relevant papers on the
subject have appeared quite recently (see
\cite{Ginzburg:2003,Ginzburg:2004,Ginzburg:2005}). In these works, the
structure of the unitarity constraints is discussed in detail within a rather
general THDM, but there is no substantial overlap with our results, since our
main point is rather a detailed explicit solution of the inequalities in
question.

   So, let us now summarize briefly our main results. We have found upper
limits for Higgs boson masses in dependence on the parameter $\nu$  that embodies an information about possible
flavour-changing neutral scalar-mediated interactions. The upper bounds are
seen to grow with increasing $\nu$ (see Tab.\ref{tab}). On the other hand, this
parameter cannot take on large values (to avoid a conflict with current
phenomenology), and thus it makes no real sense to consider the mass
estimates for an arbitrary $\nu$ ; in the present paper we restrict ourselves
to $\nu \le 0.4$ (cf. the condition used when deriving the relation \eqref{Mhsmallestbound}).
 In the case with no \CP\ violation in the scalar sector ($\xi=0$), the
relevant results are obtained from the inequalities  \eqref{Mxibound}, \eqref{MAbound}, \eqref{MHbound}, \eqref{Mhsmallbound},
and the bound for the lightest scalar is shown in eq.\eqref{Mhsmallestbound} (where one should
also pass from $\lambda'_5$ to $\lambda_5$ according to \eqref{lambdaprime}). In Section \ref{sec:numeric} we have
then verified that in the \CP-violating case these values remain the same.
The results are shown in Tab. \ref{tab}, where we have singled out the case $\nu = 0$
that corresponds to the absence of flavour-changing scalar currents. Let us
remark that in the \CP-violating case we do not distinguish between the $H$
and $A$, and in the \CP-conserving case the bounds for $H$ and $A$ are the
same.
 
    Further, we have calculated an explicit dependence of the upper limit
for the $M_h$ on the angle $\beta$ in the case with $\xi = 0$. The analytic expression
reads
 \begin{equation}
  M^2_h \le \frac{\sin^22\beta}{1+|\cos2\beta|} 
  \left( 
    \frac12 - \frac 3{32\pi}\lambda_5   
   \right)
   +\lambda_5\frac 3{16\pi}
  \label{Mhbetavysl}
 \end{equation}
(cf. \eqref{Mhbeta} with the $\lambda_5$ retrieved). The dependence of the relevant bound for
a lightest scalar boson can be obtained from eq. \eqref{Mhbetabound} and the results for
some particular values of the $\lambda_5$ are depicted in Fig.\ref{grafbeta}.

   For $\nu=0$ and $\xi = 0$, our results can be compared directly with those
published in~\cite{KKT}. We get somewhat stronger bounds for $m_A$ and $m_\pm$ since, in
addition to the set of constraints utilized in \cite{KKT}, we have employed also
the inequality $p_1<8\pi$, which stems from 
charged processes (cf. the end of Section \ref{sec:inequalities}) not considered in \cite{KKT}. On
the other hand, our estimates for $m_H$, $m_h$ and the lightest scalar coincide
with the results \cite{KKT}, since the above-mentioned extra inequality is not
used here. It is also noteworthy that the upper limits for $m_h$ and $m_H$
coincide with the SM LQT bound if they are estimated separately and,
depending on the number of the simultaneously estimated Higgs scalars, the
coefficient $1/2$ appears when we take two of them and $1/3$ when all of them
are considered. 

 In the case $\xi=0$ and $\lambda_5\ne 0$ comparison with
 \cite{AAN} is possible. Here we can compare only the corresponding numerical values,
which turn out to be approximately equal when $\lambda_5 = 0$. However, for $\lambda_5=0$
our results obviously differ from those of \cite{AAN}: in particular, the
bounds for $m_A$, $m_\pm$ displayed in \cite{AAN} appear to decrease with increasing $\lambda_5$.
The authors \cite{AAN} state that they used some fixed values of the angle $\beta$;
for the purpose of a better comparison we have therefore calculated the $\beta$-dependence
 of the upper bound for $m_h$, with the result shown in \eqref{Mhbetavysl}. As it
turns out, the $m_A$ and $m_\pm$ do not depend on $\beta$ in this case.

   Finally, let us mention that in the \CP-violating case we have not been
able to get analytic results; we have only shown, numerically, that the
maximum values of the masses in question are obtained for $\xi = 0$, i.e. the
upper mass bounds are the same as in the case with no \CP\ violation in the
scalar sector.

 \def\sand{\hline\rule[-20pt]{0pt}{50pt}}
 \def\nadp#1{\hline \multicolumn{6}{l}{ \rule{0pt}{20pt}\textbf{#1}}}

 \begin{table}[h]
   
  \centering
  \begin{tabular}{|l|c|c|c|c|c|}
    \hline
            & $\mathbf{H}$ 
            & $\mathbf{A}$ 
            & $\mathbf{H}^{\boldsymbol\pm}$ 
            & $\mathbf h$ 
            & {\bfseries lightest boson} 
   \\
    \nadp{Our results}
   \\\sand
    $m/m_\text{LQT}$ 
            & %
              \multicolumn{3}{c|}{ $\sqrt{1 + \nu \dfrac3{16\pi}}$}
            & $\sqrt{\dfrac12 + \nu\dfrac{3}{32\pi}}$ 
            & $\sqrt{\dfrac13 + \nu\dfrac{1}{12\pi}}$ 
  \\\sand 
    $m[\text{GeV}]$ 
            & %
              \multicolumn{3}{c|}{712 GeV}
            & 503 GeV  
            & 411 GeV
  \\
   \nadp{Results \cite{KKT}}
  \\\sand
   $m/m_\text{LQT}$
            & 1
            & $\sqrt{3}$
            & $\sqrt{\dfrac32}$
            & $\dfrac1{\sqrt2}$
            & $\dfrac1{\sqrt3}$
  \\\sand
   $m[\text{GeV}]$
            & 712 GeV
            & 1233 GeV
            & 872 GeV
            & 503 GeV
            & 411 GeV
  \\
   \nadp{Results \cite{AAN}}
  \\\sand
   $m[\text{GeV}]$
            & 638 GeV
            & 691 GeV
            & 695 GeV
            & 435 GeV
            & ---
  \\\hline
  \end{tabular}
  \caption{Comparison with other works}
  \label{tab}
 \end{table}

\eject

\end{document}